\documentclass[a4, showpacs, notitlepage, nofootinbib]{revtex4-1}
\usepackage{graphicx}
\usepackage[siunitx]{circuitikz}
\usepackage{amsmath}
\usepackage{filecontents, url}
\usepackage{physics}
\usepackage{braket}
\usepackage{subcaption}
\usepackage{lipsum}
\usepackage{natbib}
\usepackage{footnote}

\begin{document}

\title{A machine learning search for optimal GARCH parameters}
\author{Luke De Clerk}
\affiliation{Loughborough University, Department of Physics, Leicestershire, LE11 3TU, United Kingdom}
\author{Sergey Savel'ev}
\affiliation{Loughborough University, Department of Physics, Leicestershire, LE11 3TU, United Kingdom}
\date{\today} 

\begin{abstract}
Here, we use Machine Learning (ML) algorithms to update and improve the efficiencies of fitting GARCH model parameters to empirical data. We employ an Artificial Neural Network (ANN) to predict the parameters of these models. We present a fitting algorithm for GARCH-normal(1,1) models to predict one of the model's parameters, $\alpha_1$ and then use the analytical expressions for the fourth order standardised moment, $\Gamma_4$ and the unconditional second order moment, $\sigma^2$ to fit the other two parameters; $\beta_1$ and $\alpha_0$, respectively. The speed of fitting of the parameters and quick implementation of this approach allows for real time tracking of GARCH parameters. We further show that different inputs to the ANN namely, higher order standardised moments and the autocovariance of time series can be used for fitting model parameters using the ANN, but not always with the same level of accuracy.
\end{abstract}

\maketitle
\section{Introduction}
The ability of Machine Learning (ML) algorithms (in particular Deep Learning (DL)) to optimise processes lends itself extremely well to an application in financial data. This optimisation ability is only possible in the presence of vast amounts of data. Due to the recording of high frequency trading since the late 1980s, trading of financial securities, in particular, can be seen to lend itself well to ML algorithms. Nevertheless, this application is not solely restricted to the trading of securities as seen by the vast quantity of literature on the application of Artificial Neural Networks (ANNs) in finance, \cite{apart, deep_learning_finance, economic_forecast_nn, quant_learning, economic_variables, impact}. Particularly notably, a study undertaken by Fadlalla et al. in \cite{analysis} highlights the potential impact ANNs could have upon finance from the bond rating industry to quantifying the risk of bankruptcy. The conclusion that Fadlalla et al. reach, is that the most effective use of Artificial Intelligence (AI) in finance is in a joint approach using traditional statistical analysis alongside a ML algorithm. The reason for this conclusion and not simply a blanket use of ML algorithms is the intractability of AI outputs. ANNs can be viewed as a black box in how they reach their output. The ability of an individual to justify their decision is what allows for accountability in finance. Since the 2008 financial crash, accountability has become extremely important to reduce the prevalence of turbulent times within economic cycles. Two examples in which an AI algorithm is left unchecked, are given in \cite{ted1, ted2}. In these studies, it is seen that although the AI does complete the task, it does so in a completely different way we, as humans, would expect. It is important that users of AI understand how and why the algorithm has reached the conclusion it has. If this is not possible, there is a considerable risk for different types of bias in the method. Nevertheless, ANNs in particular are extremely effective at optimising outputs and enable the user to gain better results than would be possible via numerical analysis, \cite{white1}.

Financial data presents with a large degree of stochasticity. This is seen in the way individual decisions shape the time evolution of many financial systems, \cite{agent}. The solution to many of these problems requires the use of stochastic models, increasing the time and complexity of calculations, \cite{stochastics}. This very difficult mathematical description makes real time solutions quite challenging. Therefore, in order to make this very successful approach more applicable in the environment of high frequency trading and forecasting, we need to be able to optimise the solutions extremely efficiently. A marrying of stochastic systems and AI can bring about such optimisations and solutions much quicker than traditional techniques whilst not sacrificing accuracy.

In financial time series there are several observed facts for how the volatility of the process behaves. One such observation is the conditional nature of the heteroskedasticity of volatility. The Autoregressive Conditional Heteroskedasticity (ARCH) model, was proposed by Engle, \cite{arch_original} to account for this phenomenon in modelling of financial time series. Consequently, Bollerslev, \cite{garch_bollerslev_1}, generalised this model to extend the use to longer time histories by the presence of two stochastic processes, price and volatility. As such, the GARCH model presents a good example of an efficient stochastic description of financial data evolution. Since the inception of the Generalised Autoregressive Conditional heteroskedasticity (GARCH) models, there has been exceptional effort in the improvement of GARCH frameworks to model volatility. These improvements are driven by empirical observations of volatility and the inability of GARCH to describe several essential features of price dynamics, \cite{engle_garch, garch_alternatives, go_garch, tully, nelson_garch, figarch1, garch_optionpricing, ng_garch, nelson_garch, continuous_garch}. 

Sine the proposal of the variations of the GARCH model in response to the numerous empirical observations of price dynamics, these models have been employed to describe a plethora of different empirical datasets. One very pertinent example of GARCH models being applied to empirical data are studies with GARCH being applied alongside AI. The reason for this joint methodology is in the GARCH model's ability to capture the esoteric behaviours of volatility. In \cite{anngarch1}, Wang presents a novel hybrid AI-GARCH approach to predict option price movements of certain securities, without prior definition of a functional form of the relationship between the input variables and the option price. Moreover, Monfared et al. in \cite{hybrid_garch_2}, detail how using an ANN increases the forecast accuracy of volatility levels in different economic cycles. Here Monfared et al. use another GARCH framework and ten different time scales of the NASDAQ index's volatility, to model the future volatility of this index. However, they have to optimise their GARCH model via Maximum Likelihood Estimations before this can take place, creating a very laborious process. Another study that utilises an ANN with a GARCH framework is by Kristjanpoller et al. in \cite{ann_gold_correlations}, here a GARCH model is fed into an ANN alongside variables that directly affect the price of gold. These inputs allow for an increased accuracy of the prediction of the future level of volatility within the price of gold as much as $25\%$ of the mean percentage error value. These studies highlight the power that ANNs have within finance and also of the efficient application of GARCH alongside AI. However, these studies only concern themselves with the forecasting of financial volatility or prices. In order to use a GARCH model as efficiently as possible, the model needs to be correctly optimised to the dataset in use. In all of the studies mentioned here, this has had to be done via statistical methods. Nevertheless, if we were to use artificial intelligence to optimise the GARCH framework we employ to the dataset we wish to model, we would gain a much more accurate forecast within a reduced time span. Therefore, we present methods to fit optimised GARCH models.

We are inspired by a very prominent study in parameter fitting, undertaken by Culkin et al. in \cite{bs}. The study details an application of an ANN to the Black-Scholes option pricing formula following an approach described in \cite{lo_black}. The aim of the research is to effectively price an European call-option using the Black-Scholes equation given the inputs of this model. These include: underlying asset price, maturity dates, volatility and exercise price. Therefore, Black-Scholes formula can generate vast quantities of data, needed to train an ANN, whilst also having a defined output, the price of the call-option. This study is notable because it allows for real-time pricing of options given their input properties, allowing quicker predictions of the price of the call-option. It is seen from the work in \cite{lo_black}, the multi-layer perceptron employed achieves a very high accuracy to the analytical model's prediction. The methodology employed here has inspired us to use AI algorithms in a Generalised Autoregressive Conditional Heteroskedasticity (GARCH) model framework. Our approach will be the following; we will be using the statistical attributes (statistical moments and autocovariance of the process) of a time series as the inputs to the network, to predict the outputs, the parameters of the GARCH model. This work is necessary because of the real-time ability of ANNs to give results. The current method of fitting parameters to a GARCH model is through Maximum Likelihood Estimations and as such require intense calculations that consume large amounts of time. However, with an ANN we would be able to simply feed a number of inputs and gain the model parameters (outputs) required. Our method will be useful for the high frequency volatility measures that furthered the developments of GARCH models, \cite{realised_2, realised_3}. Work under taken in \cite{luke_1} shows that given GARCH parameter values it is easy to work out the statistical moments for the GARCH model. However, the reverse problem, finding parameter values when statistical moments are known, reduces to a set of non-linear equation that in certain circumstances become extremely time consuming to solve. It is the aim of this paper to investigate how to use an ANN to estimate GARCH parameters instead of solving non-linear equations or using Maximum Likelihood Estimation (MLE) methods.

The paper is organised as follows: in section \ref{garch}, we introduce the model that we will use to train the Artificial Neural Network upon and discuss why the ANN is needed to fit the parameters of the model. In section \ref{nn}, we introduce the mathematics behind the ANN and present the specific networks we intend to use. Section \ref{results} discusses the results of the investigation for the GARCH-normal(1,1) model and finally, section \ref{conc} concludes.

\section{Methodology: Data set creation}
\label{garch}
Generalised Autoregressive Conditional Heteroskedasticity (GARCH) models were first proposed by Bollerslev in \cite{garch_bollerslev_1} as a generalisation to Engle's earlier ARCH models, \cite{arch_original}. GARCH models use the previous variance values along with a stochastic variable, $x_t$ to forecast the future variance. The impact of these variables are determined by the parameters of the GARCH models. For a GARCH(1,1) model we have: 

\begin{equation}
\sigma_t^2 = \alpha_0 + \alpha_1 x_{t-1}^2 +  \beta_1 \sigma_{t-1}^2
\end{equation}
Where $x_t = \sigma_t Z_t$, of which $Z_t$ is an independent identically distributed random variable, with a mean of zero and variance equal to one. If we assume, the variable $Z_t$ is a conditional gaussian variable, we denote the model a GARCH-normal(1,1) model. Bollerslev in \cite{garch_bollerslev_1}, presents the analytical equations for the statistical moments for such a model. This is given by: 

\begin{equation}
E(x_t^{2m})=\frac{a_m\left[\sum_{n=0}^{m-1}a_n^{-1}(E(x_t^{2n}))\alpha_0^{m-n}{{m}\choose{m-n}} \mu(\alpha_1,\beta_1,n) \right]}{[1-\mu(\alpha_1,\beta_1,m)]}
\label{moments}
\end{equation}
where $\mu(\alpha_1,\beta_1,m)=\sum_{j=0}^{m}{{m}\choose{j}}a_i\alpha_1^j\beta_1^{m-j}$ and $a_j=\prod_{i=1}^{j}(2i-1)$.

To allow a GARCH-normal(1,1) model to describe empirical stock market data, we can, for example, fix the values of the statistical moments close to the values of empirical data. Then, we can use traditional statistics such as the Maximum Likelihood Estimators to optimise the model for the given set of moments. However, these methods become extremely difficult to implement when the equations that require solution have more than two free variables. Therefore, these methods only work for low orders of the GARCH(p,q) model, however, such methods also become very laborious and time consuming when we wish to fit parameter values in a high frequency setting. To increase the speed of parameter fitting, we can implement AI methods, more specifically, an Artificial Neural Network (ANN). For a GARCH-normal(1,1) we have the following equations for the second order moment, $E(x^2)$, the fourth order standardised moment, $\Gamma_4$ and the sixth order standardised moment, $\Gamma_6$:

\begin{equation}
E(x^2) = \sigma^2 = \frac{\alpha_0}{1-\alpha_1-\beta_1},
\label{2}
\end{equation}

\begin{equation}
\Gamma_4 = \frac{E(x^4)}{E(x^2)^2} = 3 + \frac{6\alpha_1^2}{1-3\alpha_1^2-2\alpha_1 \beta_1-\beta_1^2},
\label{4}
\end{equation}

\begin{equation}
\Gamma_6=\frac{E(x^{6})}{(E(x^{2}))^3}=\frac{15(1-\alpha_1-\beta_1)^3(1+\frac{3(\alpha_1+\beta_1)}{1-\alpha_1-\beta_1}+\frac{3(1+\frac{2(\alpha_1+\beta_1)}{1-\alpha_1-\beta_1})(\beta_1^2+2\alpha_1\beta_1+3\alpha_1^2)}{1-3\alpha_1^2-2\alpha_1\beta_1-\beta_1^2})}{1-15\alpha_1^3-9\alpha_1^2\beta_1-3\alpha_1\beta_1^2-\beta_1^3}.
\label{6}
\end{equation}
Additional to the use of higher order moments, we wish to investigate the fitting of the GARCH model parameters using the autocovariance of the time series. The reasoning for this inclusion, is due to the ability of a autocovariance calculation to include effects from a time period, rather than a specific instance in time, as is the case with moments. In \cite{garch_bollerslev_1}, Bollerslev derives an equation to determine the autocovariance with lag $n$, $\gamma_n$ of the square of returns, $cov(x^2_t, x^2_{t+n})$. The solution of this equation (see for detail appendix \ref{cov_der}) is given below.

\begin{equation}
\hat \gamma_n = \frac{\gamma_n}{E(x^2)} = \frac{2\alpha_1(1-\alpha_1\beta_1-\beta_1^2)}{1-3\alpha_1^2-2\alpha_1\beta_1-\beta_1^2}(\alpha_1 + \beta_1)^{(n-1)}
\label{coveqt}
\end{equation}
We therefore, investigate another sequence of inputs; the second order moment, the fourth order standardised moment and then the autocovariance of the squared process. For completeness, we investigate a range of lags, $n=2, 6, 10$. 

Equations (\ref{2})-(\ref{coveqt}) allow us to create several different datasets for the training of a machine learning algorithm. We use equations (\ref{2}-\ref{6}) to create a training set ($\alpha_1$; $\Gamma_4$, $\Gamma_6$, $E(x^2)$). We create another datasets to incorporate the autocovariance of the process with certain time lags, $n=2,6$ and $10$, using equations (\ref{2}), (\ref{4}) and (\ref{coveqt}), generating sets ($\alpha_1$; $\Gamma_4$, $\hat \gamma_n$, $E(x^2)$). Using these datasets we trained an ANN to predict $\alpha_1$ from the inputs of the fourth order standardised moment, $\Gamma_4$, the sixth order standardised moment, $\Gamma_6$, the second order moment, $E(x^2)$ or $\Gamma_4$, the normalised autocovariance with lag $n$, $\hat \gamma_n$, $E(x^2)$. We will then use the analytical expressions of equation (\ref{moments}) to predict the corresponding values of $\alpha_0$ and $\beta_1$. For every dataset created, we split the GARCH data into three individual sets, training data, which has $40\%$ of the overall data (50,000 data points), testing which holds $40\%$ of the data too and validating which holds $20\%$ of the data (25,000 data points) for each network trained on each model.

\section{Methodology: Artificial Neural Network Structure}
\label{nn}
Here, we employ a feedforward Artificial Neural Network (ANN) that uses a back-propagation algorithm for the minimisation of error. To do this we use the `Adam' optimiser, \cite{adam}, with a loss function of the mean squared error between the network's forecasts and the data set. All models are used within the PyTorch module of Python. 

\begin{figure}
 \centering
\includegraphics[width=\linewidth]{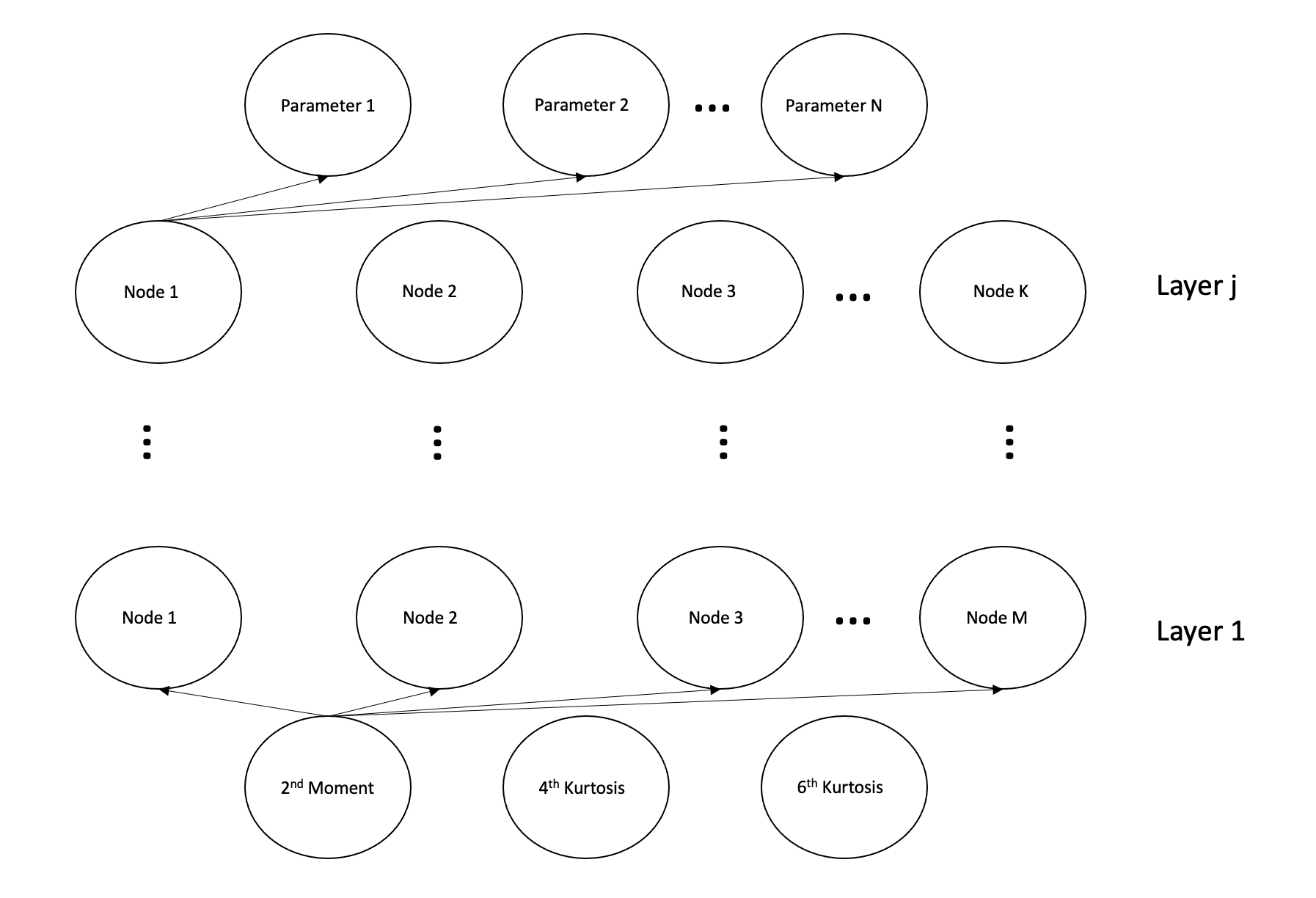}
\caption{Here, we show the schematic drawing of one of the ANN we have used. The inputs feed into the first layer of M nodes (Layer 1), with all inputs attached to each node, these then feed into subsequent layers with the final jth layer of K nodes feeding the output layer of N outputs, in our case the N number of parameters we wish to predict. All nodes in a layer are connected to all nodes in the successive layer, making it a multi-layer perceptron. In the specific ANN we use, we have four hidden layers with (128, 2048, 2048, 128) nodes in each layer, respectively.}
\label{schematic_NN}
\end{figure}

Our work follows the mathematical notation available in \cite{python_ml_book}. Here, we denote the values of each node, or the so called `Activation Unit' by $a_i^{(j)}$, where $i$ denotes the unit number (node number) and $j$ denotes the layer we are on. Very generically, we use $h$ for the learning layers (also termed hidden layers) and $out$ for the output layers. The net input to a given layer, for instance, $h$, is represented by, $z_i^{(h)}$:

\begin{equation}
z_i^{(h)} = \sum_{k=0}^{m} a_k^{(h-1)} W^{(h)}_{k, i}
\end{equation}
Where $W_{k,i}$ denotes the weight of all previous nodes connected to the present node we are analysing. In the above equation the layer number is $i$ and $a_i^{(h-1)}$ denotes the activation unit from the previous layer. For the input layer, $a^{(in)}_i$, this will be either a statistical moment or the autocovariance with specific lag $n$ of the process. These weights define the strength of connection between nodes of previous layers. Biases are used in ANNs and are denoted by $a_{k=0}^{(h-1)}$. The fundamental role of the bias is to add another parameter to tune the output of the network to minimise the loss function. The biases have a weight of one and these do not change in the network. To gain an output for this node, we pass our net input into an activation function, $\phi(.)$, as seen below. This then goes into successive connections and eventually reaches the output layer.

\begin{equation}
a_j^{(h)} = \phi(z_j^{(h)})
\end{equation}
The activation function can have different functional forms. However, it has to be differentiable to allow optimisation of the weights and biases of our network. From the analysis undertaken in \cite{activation_functions} and \cite{activation_function_comparison} by Sibi et al. and Choudhary et al., respectively. It is shown that as long as an activation function is correctly and thoroughly trained upon a data set then there is no real advantage of one over the other, hence we use the ReLU activation function. Our data is scaled to the interval $[0-1]$, the need for scaling the data in this way is to stop the network giving unnecessary biases to larger data values. In order to scale our inputs we have used the `MinMaxScale' function found in the `sklearn' module of Python. In figure \ref{schematic_NN}, we have detailed the schematic diagram for the networks we intend to use.

The values that can be tuned are termed `hyperparameters'. For the hyperparameters of our network we have chosen, to use a four layer perceptron, other networks were trialled with differing layer numbers and hyperparameters but four layers with the following parameters was shown to be the most optimal. We therefore employ our perceptron with (128, 2048, 2048, 128) nodes in each layer, respectively. Moreover, we have chosen 5000 epochs, the number of times we pass the data through the network to update the weights, and a learning rate of 0.01, the maximum magnitude the weights can be changed within one epoch. Due to the large dimensionality of the network we invoke an early stopping criteria of the network. If the network does not improve the prediction of the validation set on additional epochs, we stop the training process at the best epoch we can find. This is to stop the network overfitting the data. In essence, it prevents the network from just learning the data, which gives a good training error but is very poor at generalising the behaviour to unseen data sets. The loss function we shall employ, will be the mean squared difference (MSD) of the actual output to the predicted output of the ANN. The loss function (which we wish to minimise), for which we can track the progress of the optimisation process is detailed below:

\begin{equation}
MSD = \frac{1}{N} \sum^{N}_{i=1}(y_i-\hat y_i)^2
\label{msd}
\end{equation}
Where $y_i$ is the actual value from our training or testing data and $\hat y_i$ is the predicted value from the ANN. The outputs of our network will be the parameters of each GARCH model. For the GARCH(1,1) model, we will have one output: $\alpha_1$, the other two parameters of the GARCH(1,1) model will be calculated via the analytical expressions of the higher order moments.

\section{GARCH-normal fitting}
\label{results}
For certain orders of statistical moments to exist, we are required to set limits for the parameters, $\alpha_1$ and $\beta_1$, in the GARCH model. For $\alpha_1$ we have detailed the limits of the parameter for the different moments below. However, to ensure finiteness of the second order moments, we require the limits of $\beta_1$ be $[0, 1]$. We do not necessarily have to restrict the value of $\alpha_0$, however, to enable the ANN to train as accurately as possible on the datasets, we restrict the value of $\alpha_0$ to the interval, $[1e-03, 1e-06]$. To ensure finiteness for the $m$th order moments, we have to restrict the data to the following condition, \cite{garch_bollerslev_1}:

\begin{equation}
\sum_{j=0}^{m}{{m}\choose{j}}a_i\alpha_1^j\beta_1^{m-j} < 1
\end{equation}
For example, we have to restrict the value of $\alpha_1$ such that $0 \leq \alpha_1 \leq \frac{1}{15}^{\frac{1}{3}}$ to enable a finite value of $\Gamma_6$, \cite{garch_bollerslev_1}. Furthermore, we investigate different moment sequences namely $\sigma^2$, $\Gamma_4$ and either, $\Gamma_8$ or $\Gamma_{10}$. Therefore, we have a different limit of $\alpha_1$ to enable finiteness of the highest moment, therefore, in these examples we have $\alpha_1$ as $0 \leq \alpha_1 \leq \frac{1}{105}^{\frac{1}{4}}$ and $0 \leq \alpha_1 \leq \frac{1}{945}^{\frac{1}{5}}$, for the eighth and tenth order standardised moments, respectively.

\begin{figure}
\centering
\begin{subfigure}{.5\textwidth}
  \centering
\includegraphics[width=\linewidth]{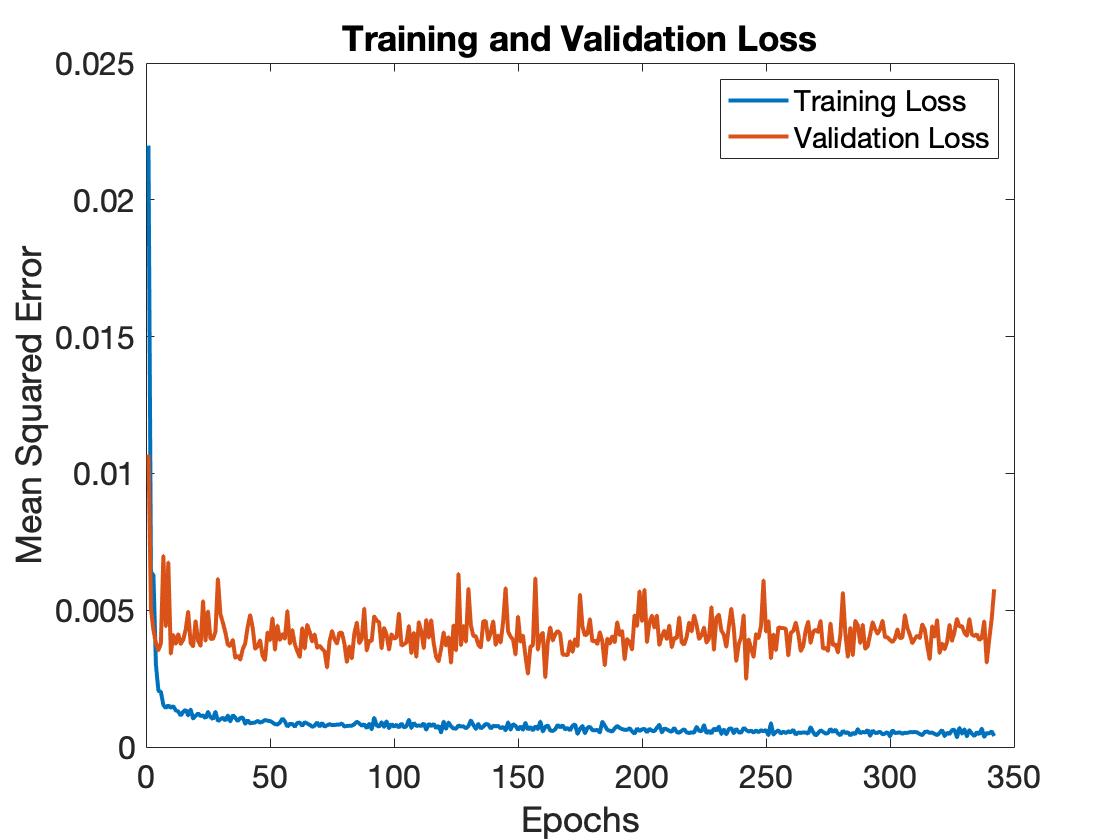} 
\caption{}
\label{garch_11_train}
\end{subfigure}%
\begin{subfigure}{.5\textwidth}
  \centering
\includegraphics[width=\linewidth]{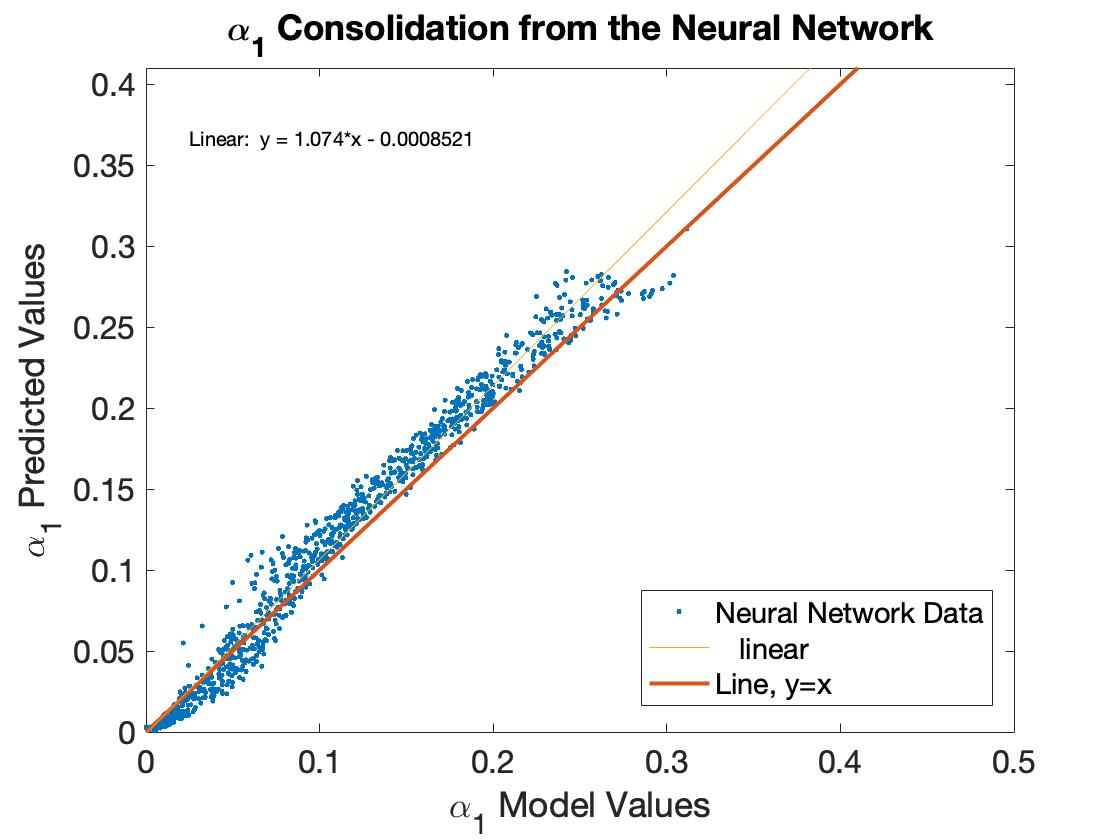}
\caption{}
\label{garch_11_test}
\end{subfigure}
\caption{In panel (a), we present the dependence of the in-sample errors for the number of epochs in the training and validating of the neural network. In panel (b), we show the out-of-sample prediction of the neural network against the actual values of $\alpha_1$ for the set of moments: $E(x^2)$, $\Gamma_4$ and $\Gamma_6$. We also indicate a perfect linear fitting in red, whilst the line of best fit is shown in yellow, approximated to the data. This gives an equation of $y=1.074x-0.001$.}
\end{figure}

In figures \ref{garch_11_train} and \ref{garch_11_test}, we see the training and validation error for the dataset ($E(x^2)$, $\Gamma_4$, $\Gamma_6$) we use and the testing results for the GARCH(1,1) parameter, $\alpha_1$. Figure \ref{garch_11_train} shows the training and validation error as a function of epoch number. The validation and training errors are independent of one another and as we would expect the more epochs we have the more accurate the network becomes. In this particular network, we have quite a noisy validation error, it is however, stable (does not wildly vary from one epoch to the next) and shows a very low error for the validation dataset, meaning we have good generalisation within the network. 

In figure \ref{garch_11_test}, we see the accuracy of the network to predict the $\alpha_1$ parameter of a GARCH(1,1). As we have had to restrict the possible values of the parameter we get an edge in parameter space around this particular boundary value. As such, if we wish to predict all three parameters of the GARCH(1,1) model it will give these edges on the corresponding limits of the parameter, reducing the prediction accuracy of the network as a whole. To increase the accuracy, we predict just one parameter with the ANN and then use:

\begin{equation}
\beta_1 = \sqrt{1-2\alpha_1^2-\frac{6\alpha_1^2}{\Gamma_{4, emp} - 3}} - \alpha_1
\label{b1_algebraic}
\end{equation}
To calculate the corresponding value of $\beta_1$, where $\Gamma_{4, emp}$, is the empirical value of the fourth order standardised moment, $\Gamma_4$ and then to calculate the value of $\alpha_0$ we use the following expression, \cite{luke_1}:

\begin{equation}
\alpha_0 = \sigma_{emp}^2(1-\alpha_1-\beta_1)
\label{a0_algebraic}
\end{equation}
Where, $\sigma^2_{emp}$ is the empirical value of the second order moment.

\begin{figure}
\centering
\begin{subfigure}{.5\linewidth}
  \centering
\includegraphics[width=\linewidth]{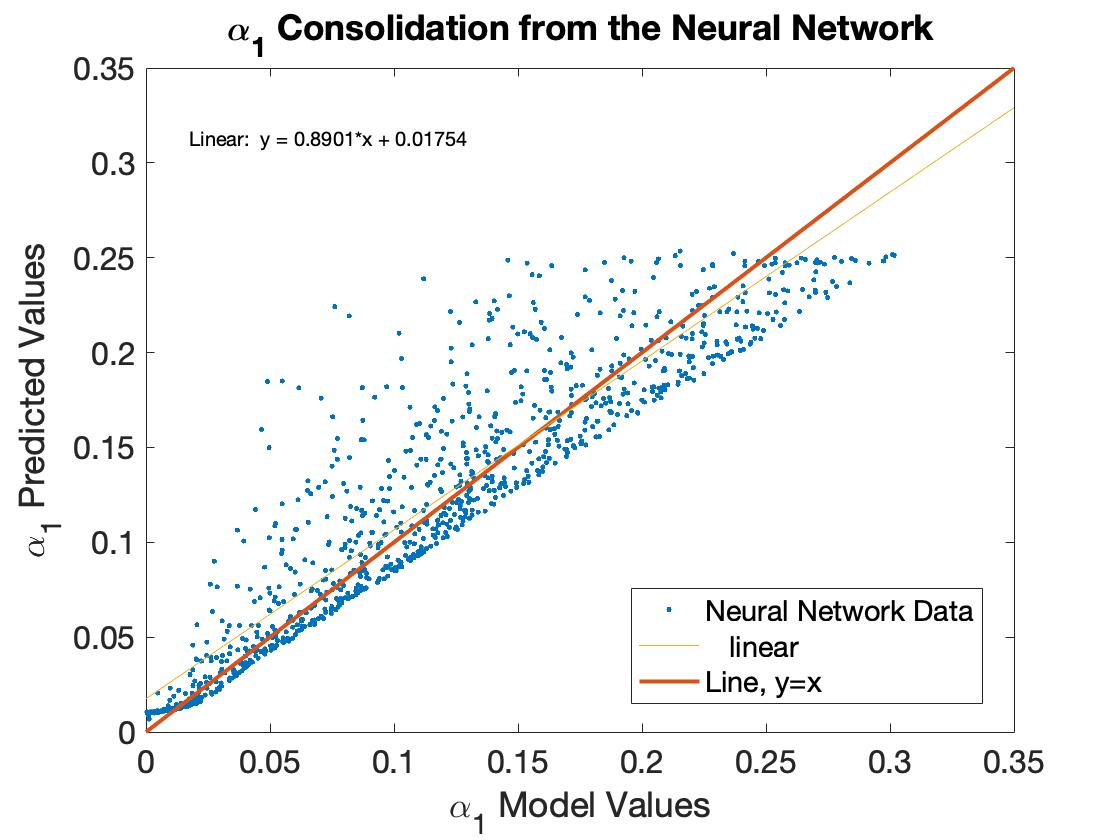} 
\caption{}
\label{g8_11}
\end{subfigure}%
\begin{subfigure}{.5\linewidth}
  \centering
\includegraphics[width=\linewidth]{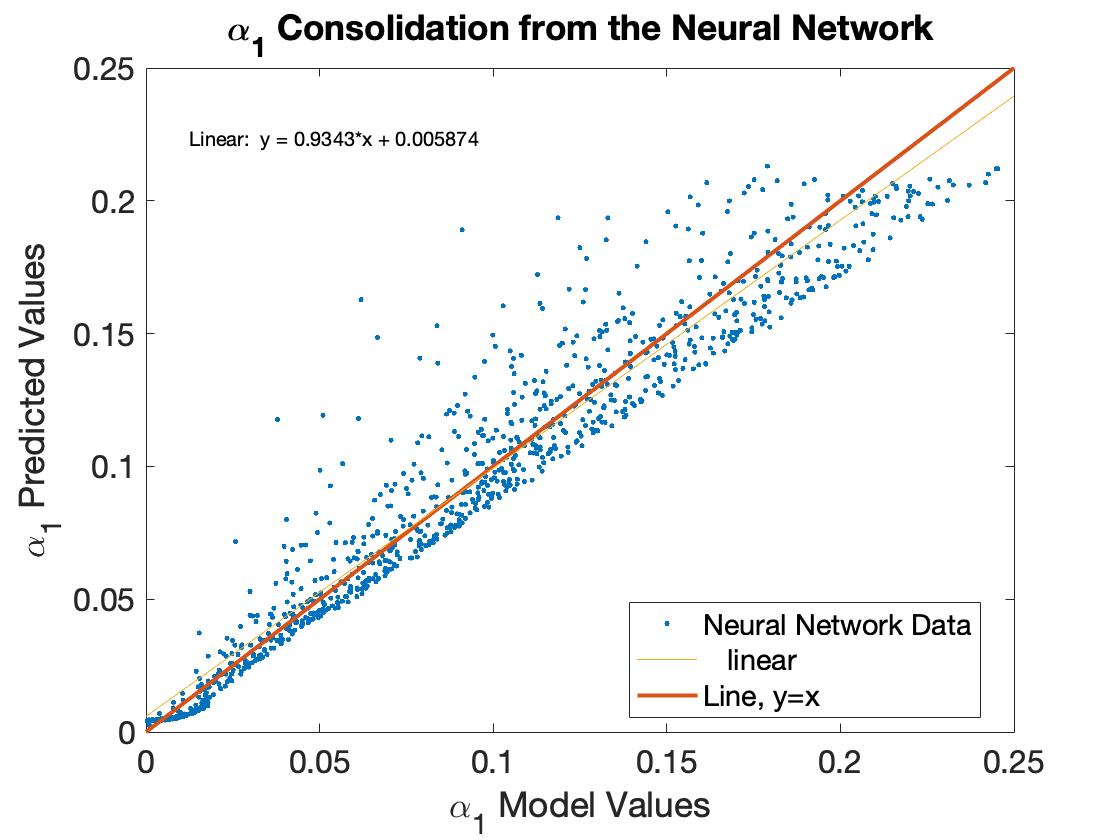}
\caption{}
\label{g10_11}
\end{subfigure}
\caption{In panel (a), we show the accuracy of the network by the prediction of the network against the actual values of the dataset when we use the eighth order standardised moment alongside the second order moment and the fourth order standardised moment as inputs to the ANN, with best fitting; $y=0.89x+0.02$, shown in yellow. In panel (b), we show the accuracy of the network as the prediction of the network against the actual values of the dataset when we use the tenth order standardised moment alongside the second order moment and the fourth order standardised moment as inputs to the ANN, with best line fitting; $y=0.934x+0.006$, shown in yellow. In both panels, we show the line $y=x$ in red to compare the fitting of the data to a perfect fitting method.}
\end{figure}

In figures, \ref{g8_11} and \ref{g10_11}, we see the $\alpha_1$ test dataset values against the predicted values of $\alpha_1$ for the networks that use $\Gamma_8$ and $\Gamma_{10}$ instead of $\Gamma_6$, respectively. When we have higher orders than the sixth standardised moment as the inputs to the network, we witness a reduction in the predictability of the GARCH parameter, $\alpha_1$. For the eighth and tenth order standardised moments, we have a reduction in the range of $\alpha_1$ values that allow these moments to exist. As a result, we witness the edge effects at a much lower value of $\alpha_1$. In the case of the tenth order standardised moment, this reduction in parameter space manifests itself as an increase in the accuracy of the network to predict $\alpha_1$. Nevertheless, in comparison to the network that uses $\Gamma_6$, figure \ref{garch_11_test}, in general there is a slight reduction in the predictive ability of the ANN to fit GARCH parameter values.

\begin{figure}
\begin{subfigure}{0.45\textwidth}
 \includegraphics[width=\linewidth]{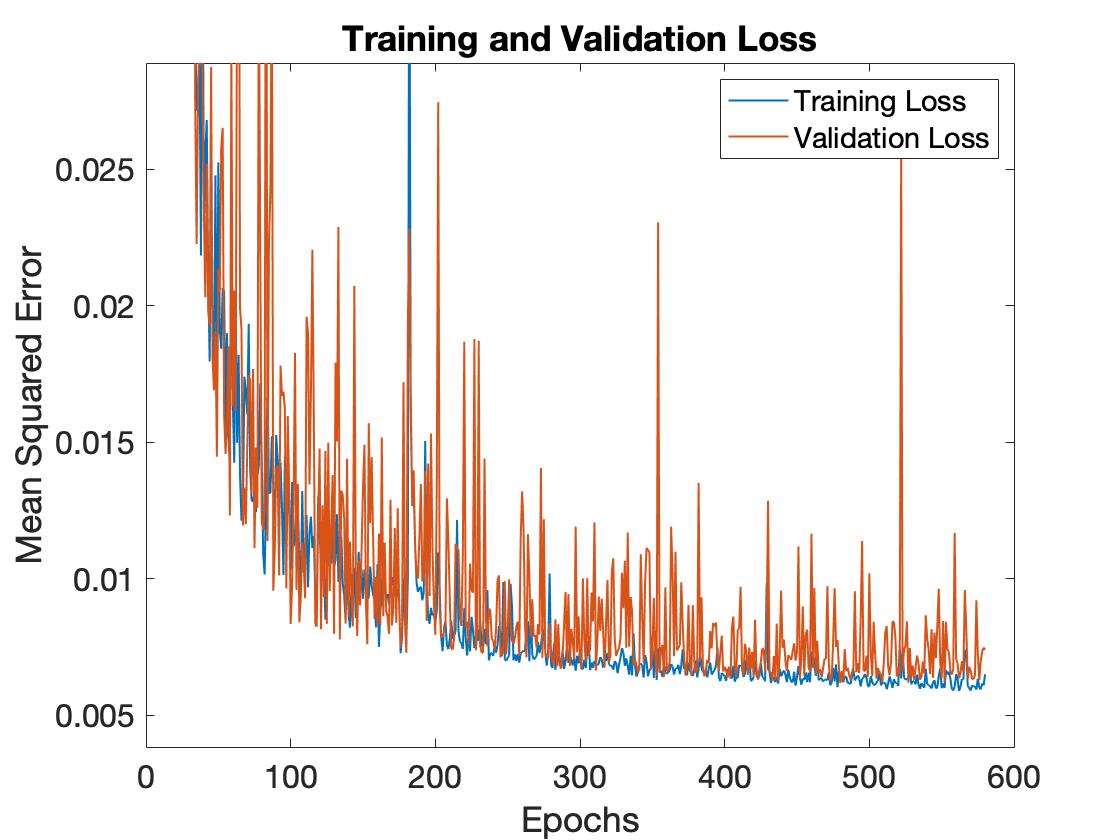}
\caption{In panel (a), we show the training and validating errors for the ANN when we use $n=2$ for the covariance alongside the second order moment and fourth order standardised moment as inputs to the ANN.} \label{cov1}
\end{subfigure}\hspace*{\fill}
\begin{subfigure}{0.45\textwidth}
\includegraphics[width=\linewidth]{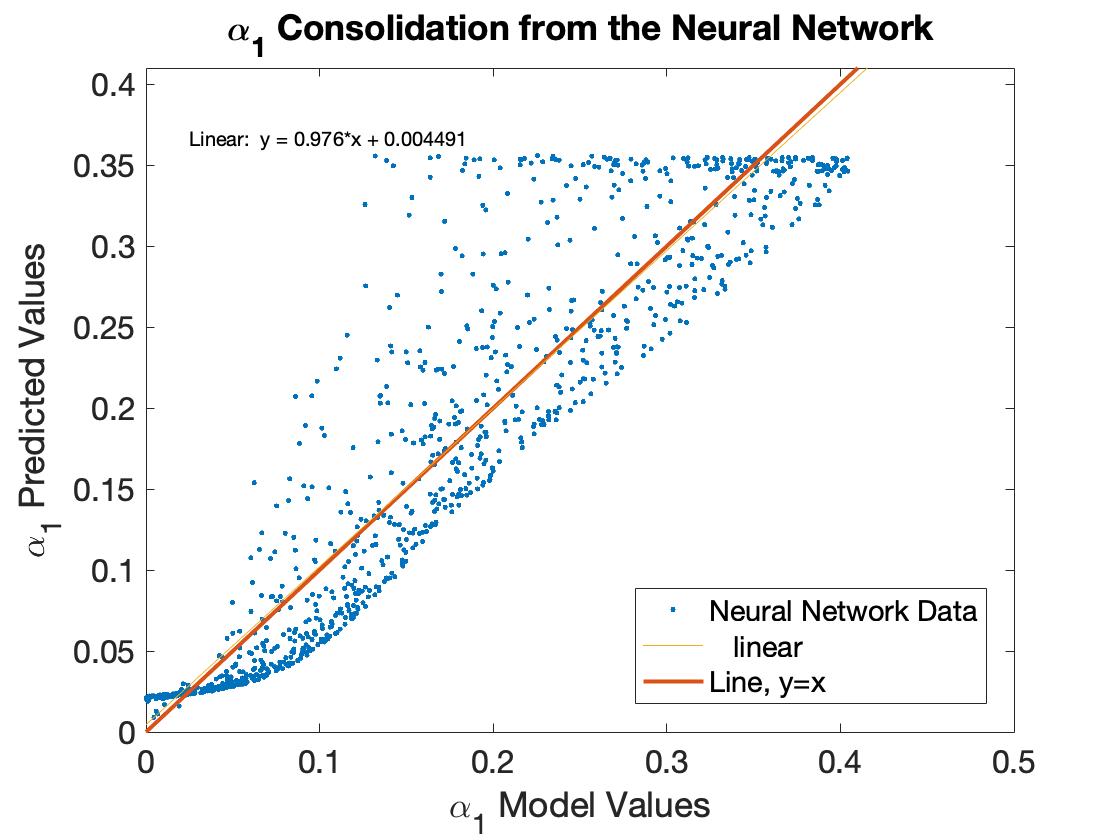} 
\caption{In panel (b), we show the out-of-sample predictions of the ANN, the linear fitting in yellow which gives a line of $y=0.98x+0.0045$ and the line $y=x$ in red.} \label{cov2}
\end{subfigure}

\medskip
\begin{subfigure}{0.45\textwidth}
  \includegraphics[width=\linewidth]{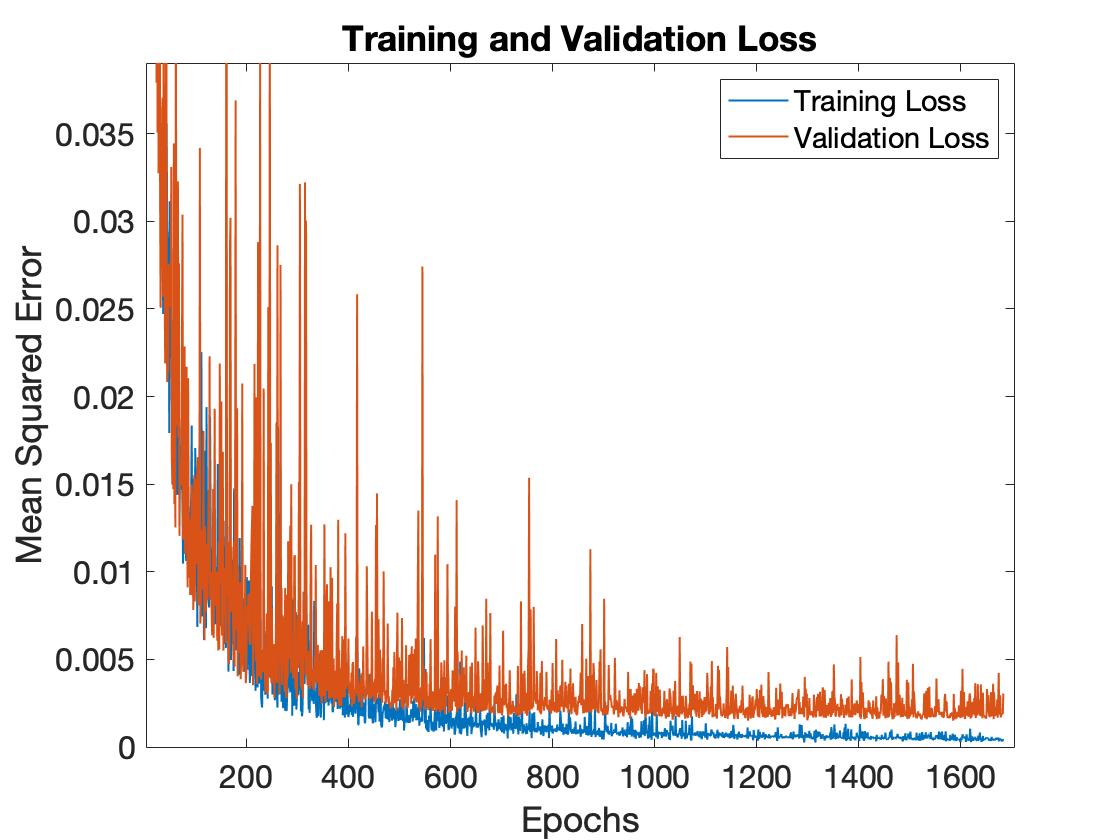}
\caption{In panel (c), we show the training and validating errors for the ANN when we use $n=6$ for the covariance alongside the second order moment and fourth order standardised moment as inputs to the ANN.} \label{c}
\end{subfigure}\hspace*{\fill}
\begin{subfigure}{0.45\textwidth}
\includegraphics[width=\linewidth]{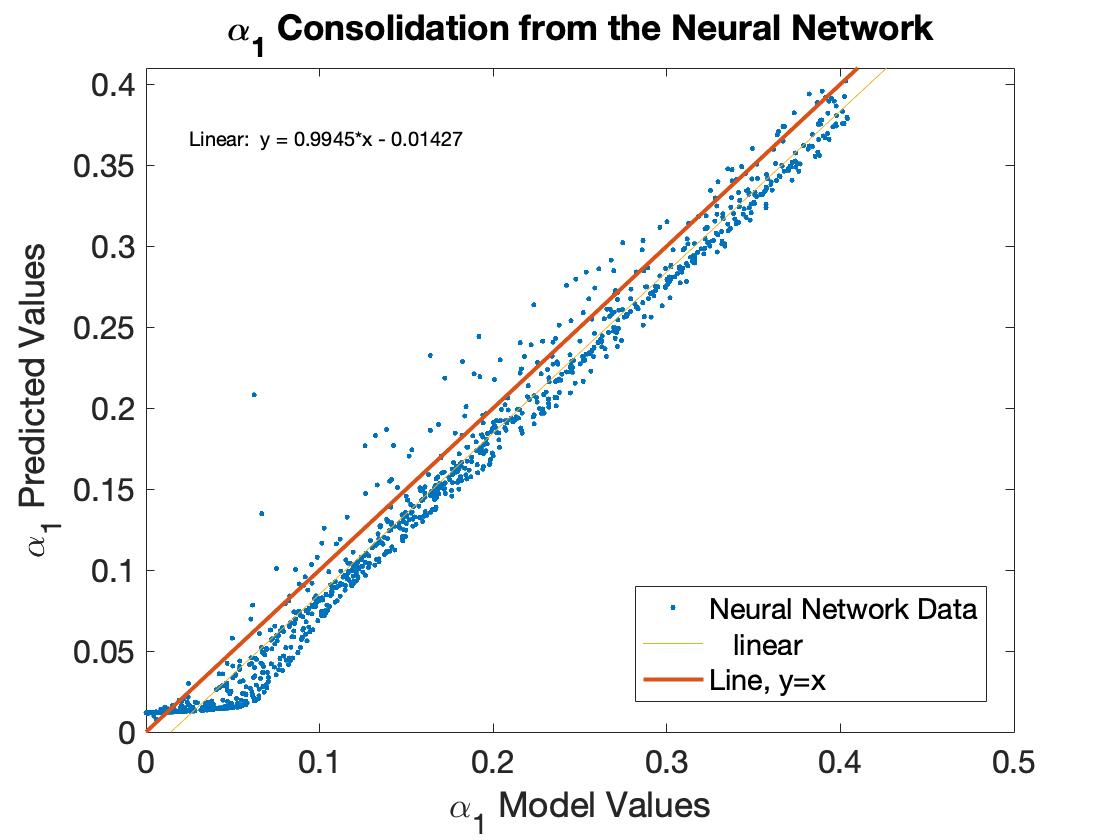} 
\caption{In panel (d), we show the out-of-sample predictions of the ANN, the linear fitting in yellow which gives a line of $y=0.99x-0.01$ and the line $y=x$ in red.} \label{d}
\end{subfigure}

\medskip
\begin{subfigure}{0.45\textwidth}
  \includegraphics[width=\linewidth]{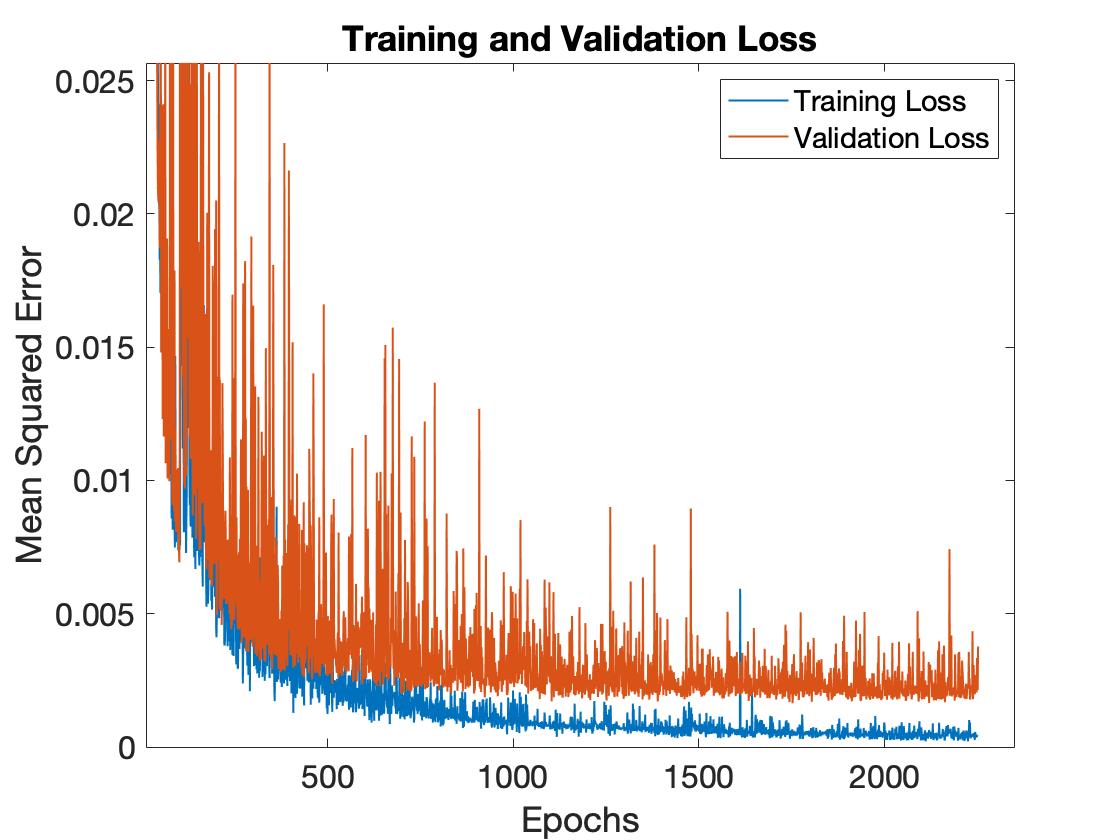}
\caption{In panel (e), we show the training and validating errors for the ANN when we use $n=10$ for the covariance alongside the second order moment and fourth order standardised moment as inputs to the ANN.} \label{e}
\end{subfigure}\hspace*{\fill}
\begin{subfigure}{0.45\textwidth}
\includegraphics[width=\linewidth]{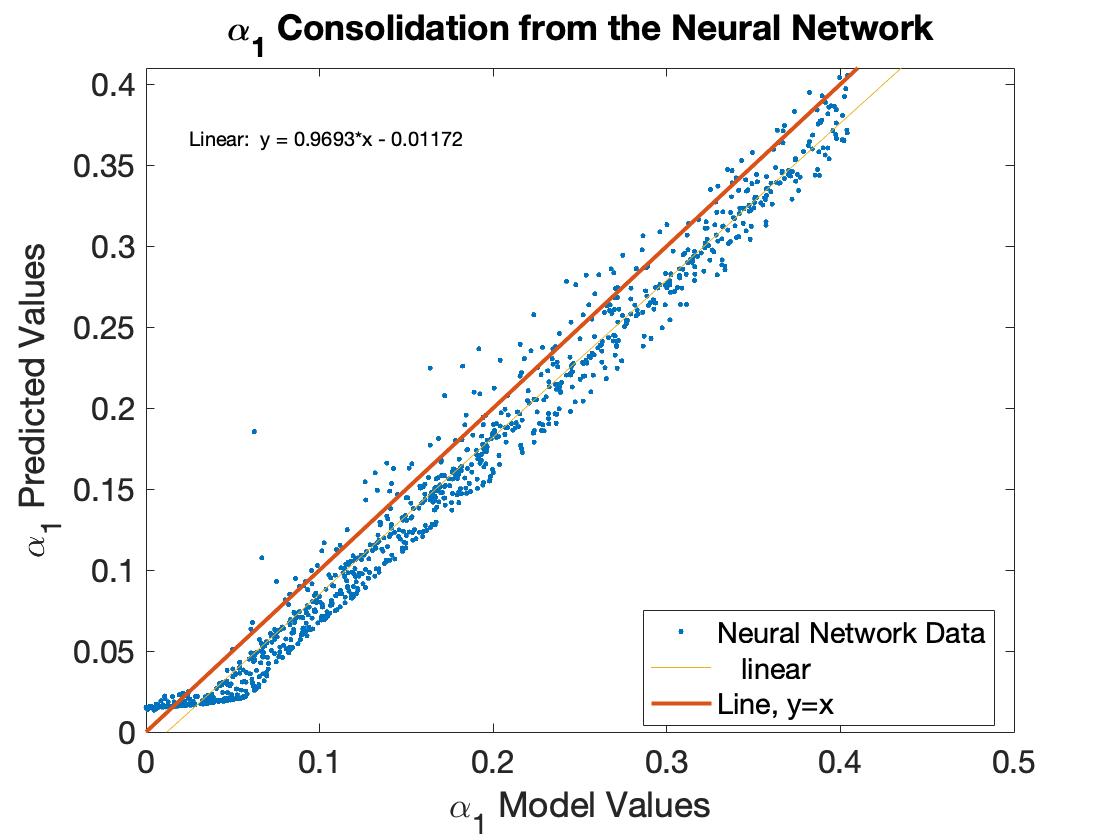} 
\caption{In panel (f), we show the out-of-sample predictions of the ANN, the linear fitting in yellow which gives a line of $y=0.97x-0.01$ and the line $y=x$ in red.}
\label{f}
\end{subfigure}

\caption{The training and validating errors against epoch number and the out-of-Sample predictions for the ANN when trained on autocovariance with different lags.} \label{long}
\end{figure}

In figures \ref{cov1} and \ref{cov2}, we see the training and validating errors when we take covariance with a lag of $n=2$ as an input of training the model and the out-of-sample predictions for $\alpha_1$, respectively. We detail the same plots for $n=6$ in figures \ref{c} and \ref{d}, respectively. Whilst for $n=10$ we see the same results in figures \ref{e} and \ref{f}, respectively. For all cases, we get a very noisy training and validating error, but due to the early stopping criteria we are able to stop the training procedure on the epoch that equates to the lowest error, this allows for good generalisation for the out-of-sample dataset. We see that there is an edge present on the upper boundary of $\alpha_1$ for a lag of $n=2$. However, for lags of $n=6$ and $n=10$, we see no edge present. Therefore, we have to conclude that these edges are an artefact of using ANNs for small lags in covariance. This observation reinforces our justification of using the ANN to predict just one parameter and then using simple relations of $\Gamma_4$, equation (\ref{b1_algebraic}) to calculate the value of $\beta_1$ and $\alpha_0$ with $\sigma^2$, equation (\ref{a0_algebraic}). From the lags investigated, we see that $n=6$ gives the best out-of-sample prediction, equating to a more accurate network for unseen data. When we compare these networks to the networks that use standardised moments instead of normalised autocovariance, we see the ANN with a lag of $n=6$, outperforms all networks we have analysed. Highlighting the most accurate network we can use is when the network has the inputs; the second order moment, the fourth order standardised moment and the normalised autocovariance with lag, $n=6$. 

\section{Conclusion}
\label{conc}
To conclude, we have introduced a novel non-parametric method of fitting GARCH model parameters via the higher order moments and the autocovariance of the process. We have shown with the example of the GARCH-normal(1,1) model that we gain high accuracy when we forecast just one parameter, $\alpha_1$, via this nonparametric approach, whilst using algebraic expressions for the remaining parameters of the GARCH-normal(1,1) model. 

The ability to use a fully trained neural network to predict the parameter values of a GARCH model increases the real-time capability of such models. Moreover, when neural networks were trained on higher order standardised moments, for instance, $\Gamma_8$ and $\Gamma_{10}$, we see a reduction in the accuracy of the fitting ability compared to the lower order standardised moment, $\Gamma_6$. When we take the normalised autocovariance, $\hat \gamma_n$, $E(x^2)$ and $\Gamma_4$ as inputs to the ANN, we see that the length of lag we take for the autocovariance over severely impacts the accuracy that we are able to predict out-of-sample data. Therefore, it is clear to see that if we wish to use autocovariance to fit parameters of a GARCH-normal model, the longer lag we can take the better, at least up to the lag of $n=6$ as this has a better out-of-sample prediction than the other lags investigated.

We have also shown that the ANN employed within the paper have the ability to generalise data extremely well. However, it has been shown that the validation and training errors are extremely noisy with respect to the increasing epoch number. Therefore, it is imperative to employ the early stopping criteria to optimise the network we train in order to not overfit the data we train tupon.

\section{Acknowledgements}
The authors would like to acknowledge the support and guidance of Professor Alistair Milne and the invaluable technical guidance of Evgeny Saveliev.

 \bibliographystyle{IEEEtran}
 \bibliography{neuralpaper}

\begin{thebibliography}{10}
\providecommand{\url}[1]{#1}
\csname url@samestyle\endcsname
\providecommand{\newblock}{\relax}
\providecommand{\bibinfo}[2]{#2}
\providecommand{\BIBentrySTDinterwordspacing}{\spaceskip=0pt\relax}
\providecommand{\BIBentryALTinterwordstretchfactor}{4}
\providecommand{\BIBentryALTinterwordspacing}{\spaceskip=\fontdimen2\font plus
\BIBentryALTinterwordstretchfactor\fontdimen3\font minus
  \fontdimen4\font\relax}
\providecommand{\BIBforeignlanguage}[2]{{%
\expandafter\ifx\csname l@#1\endcsname\relax
\typeout{** WARNING: IEEEtran.bst: No hyphenation pattern has been}%
\typeout{** loaded for the language `#1'. Using the pattern for}%
\typeout{** the default language instead.}%
\else
\language=\csname l@#1\endcsname
\fi
#2}}
\providecommand{\BIBdecl}{\relax}
\BIBdecl

\bibitem{apart}
D.~Tay and D.~Ho, ``Artificial intelligence and the mass appraisal of
  residential apartments,'' \emph{Journal of Property Valuation and
  Investment}, 1991.

\bibitem{deep_learning_finance}
J.~Heaton, N.~Polson, and J.~Witte, ``Deep learning in finance: Deep
  portfolios,'' \emph{Applied Stochastic Models in Business and Industry},
  vol.~33, no.~1, pp. 3--12, January 2017.

\bibitem{economic_forecast_nn}
F.~Aminian, E.~Suarez, M.~Aminian, and D.~Waltz, ``Forecasting economic data
  with neural networks,'' \emph{Computational Economics}, vol.~28, pp. 71--88,
  July 2006.

\bibitem{quant_learning}
J.~Spiegeleer, D.~Madan, S.~Reyners, and W.~Schoutens, ``Machine learning for
  quantitative finance: Fast derivative pricing, hedging and fitting,''
  \emph{Quantitative Finance}, vol.~18, no.~10, pp. 1635--1643, June 2018.

\bibitem{economic_variables}
S.~Thawornwong and D.~Enke, ``The adaptive selection of financial and economic
  variables for use with artificial neural networks,'' \emph{Neurocomputing},
  vol.~56, pp. 205--232, January 2004.

\bibitem{impact}
P.~Burrell and B.~Folarin, ``The impact of neural networks in finance,''
  \emph{Neural Computing and Applications}, 1997.

\bibitem{analysis}
A.~Fadlalla and C.~Lin, ``An analysis of the application of neural networks in
  finance,'' \emph{Informs}, 2001.

\bibitem{ted1}
\BIBentryALTinterwordspacing
J.~Shane. The danger of ai is weirder than you think. [Online]. Available:
  \url{https://www.ted.com/talks/janelle_shane_the_danger_of_ai_is_weirder_than_you_think}
\BIBentrySTDinterwordspacing

\bibitem{ted2}
\BIBentryALTinterwordspacing
K.~Sharma. How to keep human bias out of ai. [Online]. Available:
  \url{https://www.ted.com/talks/janelle_shane_the_danger_of_ai_is_weirder_than_you_think}
\BIBentrySTDinterwordspacing

\bibitem{white1}
Silo, ``Ai and traditional analytics- a white paper,'' Silo, Tech. Rep., 2019.

\bibitem{agent}
J.~Staccioli and M.~Napolentano, ``An agent-based model of intra-day financial
  market dynamics,'' \emph{Journal of Economic Behaviour and Organisation},
  vol. 182, pp. 331--348, 2021.

\bibitem{stochastics}
C.~Gardiner, \emph{Stochastic Methods}.\hskip 1em plus 0.5em minus 0.4em\relax
  Springer, 2010.

\bibitem{arch_original}
R.~Engle, ``Autoregressive conditional heteroscedasticity with estimates of the
  variance of uk inflation,'' \emph{Econometrica}, vol.~50, no.~4, pp.
  987--1007, July 1982.

\bibitem{garch_bollerslev_1}
T.~Bollerslev, ``Generalised autoregressive conditional heteroscedasticity,''
  \emph{The Journal of Econometrics}, vol.~31, pp. 307--327, February 1986.

\bibitem{engle_garch}
R.~Engle, ``Discussion: Stock market volatility and the crash of `87,''
  \emph{Review of Financial Studies}, vol.~3, no.~1, pp. 103--106, March 1990.

\bibitem{garch_alternatives}
T.~Bali and P.~Theodossiou, ``A conditional sgt-var approach with alternative
  garch models,'' \emph{Annals of Operations Research}, vol. 151, pp. 241--267,
  December 2006.

\bibitem{go_garch}
R.~Weide, ``A multivariate generalised orthogonal garch model,'' \emph{Journal
  of Applied Econometrics}, vol.~17, no.~5, pp. 549--564, October 2002.

\bibitem{tully}
E.~Tully and B.~Lucey, ``A power garch examination of the gold market,''
  \emph{Finance}, vol.~21, no.~2, pp. 316--325, June 2007.

\bibitem{nelson_garch}
D.~Nelson, ``Conditional heteroskedastcicity in asset returns: A new
  approach,'' \emph{Econometrica}, vol.~59, no.~2, pp. 347--370, March 1991.

\bibitem{figarch1}
R.~Baille, T.~Bollerslev, and H.~Mikkelsen, ``Fractionally integrated
  generalised autoregressive heteroskedasticity,'' \emph{Jounral of
  Econometrics}, vol.~74, no.~1, pp. 3--30, September 1996.

\bibitem{garch_optionpricing}
J.~Duan, G.~Gauthier, J.~Simonato, and C.~Sasseville, ``Approximating the
  gjr-garch and egarch option pricing models analytically,'' \emph{Journal of
  Computational Finance}, vol.~9, no.~3, April 2006.

\bibitem{ng_garch}
R.~Engle and V.~Ng, ``Measuring and testing the impact of news on volatility,''
  \emph{Journal of Finance}, vol.~48, no.~5, pp. 1749--1778, April 1993.

\bibitem{continuous_garch}
F.~Drost and B.~Werker, ``Closing the garch gap: Continuous time garch
  modelling,'' \emph{The Journal of Econometrics}, vol.~74, no.~1, pp. 31--57,
  September 1996.

\bibitem{anngarch1}
Y.~H. Wang, ``Nonlinear neural network forecasting model for stock index option
  price: Hybird gjr-garch approach,'' \emph{Expert Systems with Applications},
  vol.~36, no.~1, pp. 564--570, January 2009.

\bibitem{hybrid_garch_2}
S.~Monfared and D.~Enke, ``Volatility forecasting using a hybrid gjr-garch
  neural network model,'' \emph{Procedia Computer Science}, vol.~36, pp.
  246--253, 2014.

\bibitem{ann_gold_correlations}
W.~Kristjanpoller and M.~Minutolo, ``Gold price volatility: A forecasting
  approach using the artificial neural network-garch model,'' \emph{Expert
  Systems with Applications}, vol.~42, no.~20, pp. 7245--7521, November 2015.

\bibitem{bs}
R.~Culkin and S.~Das, ``Machine learning in finance: The case of deep learning
  for option pricing,'' \emph{Journal for Investment Management}, vol.~15,
  no.~4, pp. 92--100, August 2017.

\bibitem{lo_black}
J.~Hutchinson, A.~Lo, and T.~Poggio, ``A nonparametric approach to pricing and
  hedging derivative securities via learning networks,'' \emph{The Journal of
  Finance}, vol.~49, no.~3, pp. 851--870, 1994.

\bibitem{realised_2}
T.~Andersen, T.~Bollerslev, F.~Diebold, and P.~Labys, ``Modelling and
  forecasting reaslied volatility,'' \emph{Econometrica}, vol.~71, no.~2,
  October 2003.

\bibitem{realised_3}
T.~Andersen, T.~Bollerslev, and N.~Meddahi, ``Reaslied volatility forecasting
  and market microstructure noise,'' \emph{The Journal of Econometrics}, vol.
  160, no.~1, pp. 220--234, January 2011.

\bibitem{luke_1}
L.~D. Clerk and S.~Savel'ev, ``Non-stationary modelling of garch to fit higher
  order moments of financial series within fixed time windows,'' \emph{ArXiv:
  2102.11627}, March 2021.

\bibitem{adam}
\BIBentryALTinterwordspacing
V.~Bushaev, ``Adam- latest trends in deep learning optimization.'' [Online].
  Available:
  \url{https://towardsdatascience.com/adam-latest-trends-in-deep-learning-optimization-6be9a291375c}
\BIBentrySTDinterwordspacing

\bibitem{python_ml_book}
S.~Raschka and V.~Mirjalili, \emph{Python Machine Learning}, 2nd~ed.,
  C.~Nelson, Ed.\hskip 1em plus 0.5em minus 0.4em\relax Packt, September 2017.

\bibitem{activation_functions}
P.~Sibi, S.~A. Jones, and P.~Siddarth, ``Analysis of different activation
  functions using back propagation neural networks,'' \emph{Journal of
  Theoretical and Applied Information Technology}, vol.~47, no.~3, pp.
  1264--1268, January 2013.

\bibitem{activation_function_comparison}
A.~Choudhary, S.~Ahlawat, R.~Rishi, and V.~Dhaka, ``Performance analysis of
  feed-foreward mlp with various activation functions for handwritten numerals
  recognition,'' in \emph{2010 The 2nd International Conference on Computer and
  Automation Engineering (ICCAE)}, 2010.

\bibitem{4}
R.~Mantegna and H.~Stanley, \emph{An Introduction to Econophysics}, C.~U.
  Press, Ed.\hskip 1em plus 0.5em minus 0.4em\relax Cambridge Press, 2000, vol.
  Fourth.

\end{thebibliography}

\newpage

\appendix
\section{Covariance Derivation for GARCH(1,1) Models}
\label{cov_der}
We can write the GARCH(1,1) equation as follows, \cite{4}:

\begin{equation}
x_t^2 = \alpha_0 + (\alpha_1 + \beta_1)x^2_{t-1} - \beta_1 \nu_{t-1} + \nu_t
\end{equation}
where $\nu_t = x^2_t-\sigma^2_t$, which is serially uncorrelated with zero mean. Therefore, we can write the covariance of two distinct points in time, $cov(x^2_t, x^2_{t+n})$ as:

\begin{equation}
cov(x^2_t, x^2_{t+n}) = cov(x_t^2, (\alpha_0 + (\alpha_1+\beta_1)x_{t+n-1}^2-\beta_1\nu_{t+n-1}+\nu_{t+n}))
\end{equation}
Expanding this we can see that:

\begin{equation}
cov(x^2_t, x^2_{t+n}) = cov(x_t^2, \alpha_0) + cov(x_t^2, (\alpha_1+\beta_1)x^2_{t+n-1}) - cov(x_t^2, \beta_1\nu_{t-1}) + cov(x_t^2, \nu_{t+n})
\end{equation}
Recalling that $cov(x^2_t, x^2_{t+n}) = E[x^2_t x^2_{t+n}]-E[x^2_t]E[x^2_{t+n}]$, we can simplify the above expression to:

\begin{equation}
cov(x^2_{t+n}, x^2_{t+n}) = (\alpha_1+\beta_1)cov(x_t^2, x^2_{t+n-1}) -\beta_1(E[x_t^2\nu_{t+n-1}]-E[x^2]E[\nu_{t+n-1}])+E[x^2_t\nu_{t+n}]-E[x^2_t]E[\nu_{t+n}]
\label{cov_n}
\end{equation}
It is easy to see that given $\nu_t = x^2_t - \sigma^2_t$ and has zero mean, also, given that $x^2_t = \eta_t^2\sigma^2_t$ and that $\sigma^2_t=\sigma_t^2(\eta_{t-i})$, where $i = 1, 2, 3,...$, therefore it can be seen that $\eta_t^2$ is independent of $\sigma_t^2$. So, we can see that the above expression becomes:

\begin{equation}
\gamma_n = cov(x^2_t, x^2_{t+n}) = (\alpha_1+\beta_1)cov(x^2_t, x^2_{t+n-1})
\end{equation}
Letting, $cov(x^2_t, x^2_{t+n-1}) = \gamma_{n-1}$, we can rewrite the above into the form of a difference equation. Using the substitution: $\gamma_n = Ae^{-n\tau}$, we get the following equation:

\begin{equation}
Ae^{-n\tau} = (\alpha_1+\beta_1) Ae^{-(n-1)\tau}
\end{equation}
Dividing through by $Ae^{-(n-1)\tau}$:

\begin{equation}
e^{-\tau} = (\alpha_1+\beta_1)
\end{equation}
Therefore:

\begin{equation}
\tau = |\ln(\alpha_1+\beta_1)|
\end{equation}
We can see that due to the convergence condition for the second order moment, $\alpha_1+\beta_1 < 1$, that the above logarithm will always be negative. Resulting in the following exponential:

\begin{equation}
\gamma_n = Ae^{ln(\alpha_1+\beta_1)^n}
\end{equation}
It can be easily seen that the resulting expression for $\gamma_n$:

\begin{equation}
\gamma_n = A(\alpha_1+\beta_1)^n
\end{equation}
To solve for $A$, we recall equation (\ref{cov_n}) and we consider the case $n=1$:

\begin{equation}
cov(x^2_{t}, x^2_{t+1}) = (\alpha_1+\beta_1)cov(x_t^2, x^2_{t}) -\beta_1(E[x_t^2\nu_{t}]-E[x^2]E[\nu_{t}])+E[x^2_t\nu_{t+1}]-E[x^2_t]E[\nu_{t+1}]
\end{equation}
Substituting in for $\nu_t = x^2_t - \sigma^2_t$ and noting the zero mean of $\nu_t$:

\begin{equation}
cov(x^2_{t}, x^2_{t+1}) = (\alpha_1+\beta_1)cov(x^2_t, x^2_{t}) -\beta_1E[x^2(\eta_t^2-1)\sigma^2_t]
\end{equation}
Therefore, we get:

\begin{equation}
\gamma_1 = (\alpha_1+\beta_1)\gamma_0 -2\beta_1(E[\sigma^4_t])
\end{equation}
Replacing, $\gamma_0 = E[x^4]-E[x^2]^2$:

\begin{equation}
\gamma_{n=1} = A(\alpha_1+\beta_1) = (3\alpha_1+\beta_1)E[\sigma^4]-(\alpha_1+\beta_1)E[\sigma^2]^2
\end{equation}
Rewriting the moments as the analytical expressions of the GARCH(1,1) moments and simplifying, we gain the following equation for $A$:

\begin{equation}
A = \frac{2E[\sigma^2]\alpha_1(1-\alpha_1\beta_1-\beta_1^2)}{(\alpha_1+\beta_1)(1-3\alpha_1^2-2\alpha_1\beta_1-\beta_1^2)}
\end{equation}
Recombining our results and normalising $\gamma_n$ by the second order moment squared:

\begin{equation}
\hat \gamma_n = \frac{2\alpha_1(1-\alpha_1\beta_1-\beta_1^2)}{(1-3\alpha_1^2-2\alpha_1\beta_1-\beta_1^2)} (\alpha_1+\beta_1)^{n-1}
\end{equation}

\end{document}